\documentclass{nature}

\usepackage{color} 
\usepackage{nicefrac}
\usepackage{amsmath}
\usepackage{amssymb}
\usepackage[hypertexnames=false]{hyperref}
\usepackage{graphicx}
\usepackage[separate-uncertainty = true]{siunitx} 
\usepackage{caption}
\usepackage{sectsty} 
\usepackage{float} 

\usepackage{booktabs}

\allsectionsfont{\sffamily} 

\usepackage{braket}

\usepackage{lipsum} 

\DeclareCaptionLabelSeparator{bar}{ \textbf{$\vert$} }
\newcommand{\vbias}{V_{\mathrm{bias}}}
\newcommand{\vmod}{V_{\mathrm{mod}}}
\newcommand{\vstab}{V_{\mathrm{stab}}}
\newcommand{\istab}{I_{\mathrm{stab}}}

\definecolor{stateblue}{rgb}{0.18,0.5,.75} 
\definecolor{statered}{rgb}{0.88,0.29,.29}

\newcommand{\didu}[0]{$\mathrm{d}I / \mathrm{d}V$}

\title{Search for large topological gaps in atomic spin chains on proximitized superconducting heavy metal layers}

\author{Philip Beck$^{1,\dagger}$, Bendegúz Nyári$^{2, \dagger}$, Lucas Schneider$^1$, Levente Rózsa$^3$, András Lászlóffy$^{4}$, Krisztián Palotás$^{2,4,5}$, László Szunyogh$^{2,6}$, Balázs Ujfalussy$^{4}$, Jens Wiebe$^1$ and Roland Wiesendanger$^1$}

\begin{document}

\maketitle

\begin{affiliations}
 \item Department of Physics, University of Hamburg, Jungiusstrasse 9A, 20355 Hamburg, Germany.
 \item Department of Theoretical Physics, Institute of Physics, Budapest University of Technology and Economics, Műegyetem rkp. 3., H-1111 Budapest, Hungary.
 \item Department of Physics, University of Konstanz, D-78457 Konstanz, Germany.
 \item Wigner Research Centre for Physics, Institute for Solid State Physics and Optics, H-1525 Budapest, Hungary.
 \item ELKH-SZTE Reaction Kinetics and Surface Chemistry Research Group, University of Szeged, H-6720 Szeged, Hungary.
 \item ELKH-BME Condensed Matter Research Group, Budapest University of Technology and Economics, Műegyetem rkp. 3., H-1111 Budapest, Hungary. 
 \newline
 $\dagger$  These authors contributed equally to this work.
\end{affiliations}

\begin{abstract}
\section*{Abstract}
One-dimensional systems comprising $s$-wave superconductivity with meticulously tuned magnetism and spin-orbit coupling can realize topologically gapped superconductors hosting Majorana edge modes whose stability is determined by the gap's size. The ongoing quest for larger topological gaps evolved into a material science issue. However, for atomic spin chains on superconductor surfaces, the effect of the substrate's spin-orbit coupling on the system's topological gap size is largely unexplored. 
Here, we introduce an atomic layer of the heavy metal Au on Nb(110) which combines strong spin-orbit coupling and a large superconducting gap with a high crystallographic quality enabling the assembly of defect-free Fe chains using a scanning tunneling microscope tip. Scanning tunneling spectroscopy experiments and density functional theory calculations
reveal ferromagnetic coupling and ungapped YSR bands in the chain despite of the heavy substrate. By artificially imposing a spin spiral state our calculations indicate a minigap opening and zero-energy edge state formation. The presented methodology paves the way towards  a material screening of heavy metal layers on elemental superconductors for ideal systems hosting Majorana edge modes protected by large topological gaps. 
\end{abstract}


\section*{Main}
Inducing spin-orbit coupling (SOC) in nanostructures has recently been of great interest in a variety of  disciplines related to surface science\cite{Soumyanarayanan2016} due to its close ties with the existence of non-collinear magnetic states\cite{Dzyaloshinsky1958, Moriya1960,Bode2007}, spin-split surface states\cite{Lashell1996,Ast2007}, topological surface states\cite{Fu2007, Hasan2010} and topological superconductivity\cite{Fu2008, Lutchyn2010, Potter2012} which can be accompanied by Majorana bound states (MBS)\cite{Mourik2012, Kjaergaard2012, Pientka2013, NadjPerge2013, NadjPerge2014, Ruby2015, Kim2018}. Since the latter is a promising candidate as a building block of 
topological quantum computation\cite{Freedman2002}, systems which may potentially host MBS have attracted a lot of interest. MBS may be realized in chains of magnetic atoms, also called atomic spin chains, on $s$-wave superconductors\cite{Pientka2013, Kim2018, Schneider2020, Pientka2015}, artificially fabricated by atom manipulation with the tip of a scanning tunneling microscope (STM)\cite{Eigler1990}. Although first experimental realizations, e.g. Fe chains on Re(0001), show signatures of MBS in scanning tunneling spectroscopy (STS)\cite{Kim2018}, the system suffers from the small energy gap $\varDelta_{\textrm{s}}$ of the superconducting rhenium substrate, making a clear allocation of in-gap features difficult\cite{Schneider2020}. More recent results of such atomic spin chains on Nb(110)\cite{Schneider2021a, Schneider2021b, Kuster2021a} circumvent this, but presumably at the price of lower SOC, which manifests itself in the dominance of collinear magnetic ground states due to weak Dzyaloshinskii--Moriya interaction (DMI) terms\cite{Beck2021}, and the hybridization of the spatially extended precursors of MBS in experimentally accessible chain lengths\cite{Schneider2021b}.
\newline
Aiming at maintaining the largest $\varDelta_{\textrm{s}}=\SI{1.50}{\milli \electronvolt}$ of all elemental superconductors combined with the possibility of atom manipulation which the Nb substrate offer, there are two apparent approaches to induce a higher SOC in the system.
On the one hand, one may think of using atoms with larger atomic SOC, e.g. rare earth metals, for the formation of the atomic spin chain.
On the other hand, one can try to couple the chain to a heavy metal substrate with potentially larger SOC, which is grown on Nb(110) and therefore becomes superconducting by proximity. First attempts combining both approaches using Gd atoms on bismuth thin films grown on Nb(110) show hybridization of the Yu--Shiba--Rusinov (YSR) states of small ensembles of 3-4 Gd atoms\cite{Ding2021}. However, the assembly of longer chains and a formation of bands from the hybridizing YSR states, which are both prerequisites for the emergence of topological superconductivity and MBSs, were not possible for that system. Moreover, it is even unclear, whether a larger SOC in the constituents, i.e. chain, substrate, or both of them, will lead to a larger SOC in the YSR bands of the hybrid chain system which is finally relevant for the emergence of topological superconductivity with a large topological gap and well-defined MBSs isolated at the ends of the chain.
\newline
Here, we investigate these questions pursuing the second approach by constructing Fe chains on ultrathin Au films grown on Nb(110) used as a substrate. Au is well known to exhibit large SOC\cite{Lashell1996, Henk2004} and the proximity to Au has been demonstrated to enhance SOC-induced effects in light elements, including the scattering rate\cite{Bergman1982}, the Rashba spin-splitting\cite{BIHLMAYER2006,Marchenko2012}, and also the magnetocrystalline anisotropy energy\cite{Szunyogh1995}. Moreover, twisted spin textures were predicted to occur around single Fe atoms\cite{Lounis2012} and in Mn chains\cite{Cardias2016} on Au(111). Since previous LEED studies hint at the possibility to grow pseudomorphic thin films of Au on Nb(110)\cite{Ruckman1988}, it is a natural candidate to be used as a proximitized superconducting heavy metal layer on Nb(110).
Combining experimental STS and density functional theory (DFT) calculations by solving the fully relativistic Dirac--Bogoliubov--de Gennes (DBdG) equations of the single Fe adatom, dimers and chains, we study (i) whether there is indeed a topological gap opening in the YSR bands due to the large SOC in the Au substrate, and (ii) the effect of a different spin structure in the chain on the topological gap width and the localization of MBSs.

\subsection*{Monolayer Au on Nb(110)}

We first describe the growth and the superconducting properties of the heavy metal layer. We aimed at the preparation of ultrathin Au films, maintaining the surface structure of Nb(110) as it offers multiple distinct building directions for artificial chains, which enables some tuning of the hybridization of YSR states\cite{Beck2021, Kuester2021} and, therefore, of the in-gap band structure of such chains\cite{Schneider2021a, Schneider2021b}. An overview STM image of the Au/Nb(110) sample after the low-temperature deposition of Fe atoms is shown in \autoref{fig1}a (preparation details in the Methods section). The Nb(110) surface is almost completely covered by one monolayer (ML) of Au (see sketch in bottom panel of \autoref{fig1}a), with only a few remaining holes. Pseudomorphic growth, as predicted by LEED studies\cite{Ruckman1988}, is confirmed by manipulated atom STM images\cite{Stroscio2004} (Supplementary Note 1 and Supplementary Figure 1) of the first ML. Partially, the second ML Au has started to grow at step edges and in the form of a few free-standing islands. In this ultrathin limit, we find that the energy gap of the superconducting Nb is fully preserved in the ML Au due to the proximity effect, as demonstrated by the deconvoluted \didu~spectrum (see Methods and Supplementary Note 2 for the deconvolution procedure) taken on the bare ML Au on Nb(110) far from any Fe atom in \autoref{fig1}c (gray curve). The energy gap on the ML Au is of equal size as that of bare Nb(110) at our measurement temperature and the in-gap \didu~signal is zero (see Supplementary Figure 2). This behavior is crucial for our experiments, but might be altered for thicker films as indicated by recent theoretical\cite{Csire2016, Csire2016a} and experimental\cite{Gupta2004, Tomanic2016} studies.

\subsection*{YSR states of single Fe atoms}
We continue with the investigation of the YSR states induced by single Fe atoms, i.e. the building blocks of chains. According to our DFT calculations the Fe adatom has a magnetic moment of $3.57 \mu_\textrm{B}$ with a preferred orientation perpendicular to the Au surface (Supplementary Note 5). A close-up STM image of the statistically distributed single Fe adatoms is shown in \autoref{fig1}b, where they appear as shallower protrusions on the first ML and as brighter spheres on the second ML. We restrict ourselves to the first ML since only this layer has large enough terraces to construct artificial chains. The Fe adatoms are adsorbed on the fourfold coordinated hollow sites on this ML Au (Supplementary Note 1 and Supplementary Figure 1). Fe adatoms which are adsorbed far from other adatoms or defects show similar \didu~spectra as shown in the top panel of \autoref{fig1}c (red curve). This reproducibility is required for a well-defined band formation in bottom-up fabricated nanostructures made from such individual adatoms. On the adatoms we find two pairs of YSR states induced in the gap of the Au ML which are marked by black arrows and greek letters\cite{Yazdani1997,Ruby2016, Choi2017}. One of them is energetically located close to the Au ML gap edge ($\pm \alpha$, $\pm \SI{1.23}{\milli \electronvolt}$) while the other one is located close to the Fermi energy $E_\mathrm{F}$ ($\pm \beta$, $\pm \SI{0.27}{\milli \electronvolt}$). We use constant-contour maps (see Methods) to resolve the spatial distribution of both YSR states\cite{Ruby2016, Choi2017}, as shown in \autoref{fig1}d.
The $\pm \alpha$ state has a spatial distribution resembling that of a $d_{yz}$ orbital with two lobes pointing along the $[1\overline{1}0]$ direction. On the other hand, a spatial distribution resembling that of a $d_{xz}$ orbital extended along the $[001]$ direction is observed for the $\pm \beta$ state. Note that the shapes of the $+\beta$ and the $-\beta$ state are somewhat different since the former has additional faint lobes along the $[1\overline{1}0]$ direction probably indicating contributions from a $d_{x^2-y^2}$-like YSR state.
The spatial distributions are explainable by the orbital origin of YSR states\cite{Ruby2016, Choi2017,Beck2021} and the point group $C_{2\textrm{v}}$ for the system. However, note that the energetic order of the states is interchanged with respect to the case of bare Nb(110)\cite{Beck2021}, and that there are no obvious indications for the other two possible, $d_{xy}$- and $d_{z^2}$-like, YSR states, which might indicate an overlap of these peaks in energy with the $d_{xz}$-, $d_{yz}$-, or $d_{x^2-y^2}$-like YSR states, or that they are hidden in the coherence peaks of the substrate.
\newline
In order to further clarify these experimental results we calculated the local density of states (LDOS) for a single Fe atom on the Au/Nb(110) film as shown in the top panel of \autoref{fig2}a and \autoref{fig2}b (see Methods for the calculation details). There are three very close-by, almost overlapping YSR states in the vicinity of the substrate's coherence peaks (see also Supplementary Figure 5a). They correspond to the $d_{z^2}$, $d_{xy}$ and $d_{yz}$ orbitals (see the $+1.32$ meV map in \autoref{fig2}b) and the spatial distributions of the LDOS around these peaks are very sensitive to the exact energy (Supplementary Figure 5b). Due to their energetic location and orbital symmetries, we assign them to the experimental $\pm\alpha$ YSR state (c.f. \autoref{fig1}c and d) which appear as a single peak in the \didu~spectrum due to the finite-temperature smearing. Additionally, there are two most intense peaks in the calculations which stem from two energetically close-by YSR states near $E_\mathrm{F}$ where the one closest to $E_\mathrm{F}$ corresponds to the $d_{x^2-y^2}$ orbital (see the $+0.38$ meV map in \autoref{fig2}b), and the less intense one further apart from $E_\mathrm{F}$ to the $d_{xz}$ orbital (see the $-0.58$ meV map in \autoref{fig2}b). While the latter resembles the experimentally observed $-\beta$ state, the former has more similarities to the $+\beta$ state (c.f. \autoref{fig1}d). These theoretical results indicate that the different spatial experimental distributions of the $+\beta$ and $-\beta$ states in \autoref{fig1}d can be explained by supposing that they correspond to two different YSR states of $d_{x^2-y^2}$ and $d_{xz}$ orbital character which overlap within the experimental energy resolution, rather than to a single YSR state. The two YSR peaks also overlap in the theoretical calculations if a larger imaginary part is chosen for the energy, corresponding to a higher effective temperature. Finally, note that the electron-hole asymmetries in the intensities of the calculated peaks appear to be inverted compared to the experiment. With the exception of the $d_{xz}$-like YSR state, each pair of peaks has a larger electron contribution above $E_\mathrm{F}$ (Supplementary Figure 5a). The $d_{xz}$-like YSR state has a higher electron contribution below $E_\mathrm{F}$ which implies that this state has the strongest coupling to the substrate.

\subsection*{YSR states of Fe dimers}
Before we continue with the investigation of Fe dimers, we consider some intuitive ideas about the most promising orientations of chains built from individual Fe atoms towards the goal of topologically gapped YSR bands. As found in previous works\cite{Schneider2021a, Schneider2021b, Kuster2021a,Liebhaber2022}, enabling a sufficient hybridization of a YSR state which is already close to $E_\mathrm{F}$, while, at the same time, minimizing the hybridizations of all the other YSR states far from $E_\mathrm{F}$ may lead to a single YSR band overlapping with $E_\mathrm{F}$. Together with SOC, this can be a sufficient condition for the opening of a topologically non-trivial gap in the lowest-energy band. Starting from the experimentally detected shapes and energies of the $\alpha$ and $\beta$ YSR states (\autoref{fig1}d) we thus regard chains along the $[001]$ direction as most promising. For this orientation, we expect weak and strong hybridizations, respectively, for the $\alpha$ and $\beta$ YSR states which are far and close to $E_\mathrm{F}$. While a manipulation of close-packed dimers along $[001]$ turned out to be impossible, we were able to tune the system into the above  conditions using dimers with a distance of $2a$ along $[001]$ (see STM image in \autoref{fig1}e). A \didu~spectrum measured above the center of the dimer as well as constant-contour maps of the spatial distributions of the three evident states are displayed in the bottom panel (orange curve) of \autoref{fig1}c and \autoref{fig1}e, respectively. In this configuration, the $\pm \alpha$ YSR states of the two atoms do not overlap significantly such that they do not split into hybridized states, but only slightly shift in energy. In contrast, the $\pm \beta$ YSR states of the two atoms strongly overlap, and split into an energetically higher one with a clear nodal line in the center between both impurities ($\pm \beta_\textrm{a}$) and another energetically lower one with an increased intensity in the center ($\pm \beta_\textrm{s}$).
\newline
These experimental conclusions are corroborated by our calculations (bottom panel of \autoref{fig2}a and \autoref{fig2}c). Apparently, all five pairs of single-atom YSR states are split, as expected from previous experimental and theoretical studies\cite{Beck2021,Nyari2021}. Although based on the orbital decomposition it is possible to separate all of the ten pairs (Supplementary Figure 6), the splitting of the three YSR states contributing to the $\alpha$ YSR state is particularly small, in accordance with the experiment, which makes it hard to resolve them in the total LDOS. In \autoref{fig2}c we plot the LDOS maps of the six most relevant peaks in \autoref{fig2}a. We find that the very weakly splitted $d_{yz}$ YSR states at $+1.32$ meV and $+1.30$ meV which are assigned to the experimental $\alpha$ YSR state appear with an almost identical shape as the single atom $d_{yz}$ YSR state (c.f. \autoref{fig2}b). In contrast, the $d_{x^2-y^2}$ and $d_{xz}$ YSR states strongly split into states with larger (at $+0.79$ meV and $-0.53$ meV) and smaller (at $+0.04$ meV and $-0.66$ meV) intensities in the center between both impurities and are thus associated with the experimental $\pm\beta_{s}$ and $\pm\beta_{a}$ YSR states, respectively. We thus conclude, that while the $\alpha$ YSR states hybridize only very weakly, the $\beta$ YSR states hybridize strongly and split into states which resemble anti-symmetric and symmetric linear combinations of the single atom YSR states\cite{Ruby2018, Flatte2000, Morr2003}, which can be seen as a prerequisite for band formation from the hybridizing $\beta$ YSR states.
\newline
\subsection*{Gapless YSR band in ferromagnetic Fe chains on Au monolayer}
Having identified a promising orientation and interatomic spacing from the investigation of the single atom and the dimer above, we move on to study artificial chains with the same interatomic separation, called  $2a-[001]$ chains in the following. A sketch illustrating this geometry and an STM image of a nine Fe atoms long $\mathrm{Fe}_9$ $2a-[001]$ chain are shown in the top panels of \autoref{fig3}a and b. Spin-polarized measurements of a $\mathrm{Fe}_{19}$ $2a-[001]$ chain indicate that the atoms in this chain configuration prefer ferromagnetic alignment (Supplementary Note 3). This is also supported by our DFT calculations (Supplementary Note 5). We found that the DMI is around 10\% of the Heisenberg exchange interaction in the dimer. Although this is not particularly weak, the SOC in the Au layer additionally induces a very strong out-of-plane on-site anisotropy, which prevents the formation of spin-spirals and stabilizes a normal-to-plane ferromagnetic spin structure.
\newline
A \didu~line profile (see Methods) was measured in the center of such a chain along its main axis and is plotted in \autoref{fig3}a (bottom panel) alongside the acquired stabilization height profile (middle panel). The first apparent characteristic of this measurement is the modulation of every feature with the interatomic spacing of $2a$ in these chains, which is also visible in the height profile. It should be emphasized that this is not a feature of the chains' in-gap band structure but is just due to the lattice-periodic part of the wave function. However, we find additional states with different well-defined numbers of maxima at increasing energy and also very close to $E_\mathrm{F}$ as indicated by the labels $n_\beta$ ($n_\beta-1$) for the numbers of maxima (nodes). Note that all these states have particle-hole partners occurring on the other side of $E_\mathrm{F}$ with the same energetic distance to $E_\mathrm{F}$ and equal numbers of maxima and nodes. However, they mostly have much smaller intensities such that they are barely visible. These pairs of states can thus be assigned to confined Bogoliubov-de-Gennes (BdG) quasiparticles residing in a YSR band induced by the finite magnetic chain in the superconductor\cite{Schneider2021a}.
To determine the orbital origin of these states, we show \didu~maps (see Methods) of the $\mathrm{Fe}_9$ $2a-[001]$ chain in \autoref{fig3}b. We find that the confined BdG states identified before in \autoref{fig3}a are localized inside the spatial extent of the chain deduced from the STM image (dashed red elliptical circumference). We assign those states to a band formed by the strong hybridization of the $\pm\beta$ YSR states of the single adatoms as they are expected to be largely localized along the longitudinal axis of the chain. Additionally, there is a state at a similar energy as the single adatom and dimer $\pm\alpha$ YSR states around $\pm \SI{1.09}{\milli \electronvolt}$. This state has exactly as many maxima as there are atoms in the chain, namely 9, which are spatially localized along both sides of the chain with a similar distance to the chain axis as the lobes of the single adatom and dimer's $\pm\alpha$ YSR states (c.f. \autoref{fig1}d and e). Therefore, we assign this state to the very weakly hybridizing $\pm\alpha$ YSR states of the single atom. The state is not observed in the \didu~line profile of \autoref{fig3}a due to its nodal line along the longitudinal chain axis.
\newline
In order to measure the dispersion of the confined BdG states from the $\beta$ YSR band, we collect similar \didu~line profiles as the one in \autoref{fig3}a of defect-free chains for lengths ranging from $N = 7$ to $N = 14$ atoms ($\mathrm{Fe}_7 - \mathrm{Fe}_{14}$, see Supplementary Note 4 and Supplementary Figure 4). It can be observed that the confined BdG quasiparticle states shift in energy as a function of the length $L = N \cdot d = N \cdot 2\mathrm{a}$ of the chain, as expected from the length-dependent interference condition
\begin{equation}
    \label{eq:interference_condition}
    q = |\textbf{q}| =\pm \frac{2 \pi n}{L}
\end{equation}
where $n$ is an integer and $|\textbf{q}|$ is the length of the BdG quasiparticle scattering vector\cite{Schneider2021a}. For particular chain lengths, the confined BdG quasiparticle states can be located very close to $E_\mathrm{F}$ (c.f. $\mathrm{Fe}_8$ and $\mathrm{Fe}_{10}$ in Supplementary Figure 4).
We perform one-dimensional fast Fourier transforms (1D-FFT) of the columns of the \didu~line profiles at fixed energy $E$ averaging all data sets taken for chains of multiple lengths, and thereby obtain the dispersion of the scattering vectors $E(q)$ (\autoref{fig3}c). This dispersion is closely linked to the $\beta$ YSR band structure. We find that this band has an approximately parabolic dispersion ranging from $\SI{-0.9}{\milli \electronvolt}$ at $q/2=0$ to $+\SI{0.5}{\milli \electronvolt}$ at $q/2=\pi / d$. Note that, as already discussed for the \didu~line profiles above, the particle-hole partner of this band has a much lower intensity. It is only visible around the Brillouin zone center ($q/2=0$) in our measurements. Most importantly, the $\beta$ YSR band smoothly crosses $E_\mathrm{F}$ without any indications of a minigap opening.
\newline
An overall similar behaviour is found using our \textit{ab-initio} framework. We performed calculations for $2a-[001]$ chains of lengths ranging from 9 to 19 Fe atoms with ferromagnetic spin alignment (Supplementary Note 7 and Supplementary Figure 7). Exemplarily, the calculated LDOS along a Fe$_9$ chain is shown in \autoref{fig4}a and can be directly compared to the measured line profile in \autoref{fig3}a. The band formation of the YSR states can clearly be observed in a wide range of the substrate gap in the form of LDOS lines with a well-defined number of maxima along the chain as indicated in the figure. In \autoref{fig4}b we present the corresponding spatial distributions of the LDOS of the Fe$_9$ chain in the form of two-dimensional maps for a selection of confined BdG states with the indicated dominant orbital characters and numbers of maxima (see the Methods section for calculation details). The states closest to the substrate's coherence peaks with $n_{yz}=2$ (and admixed $n_{yz}=4$) and $n_{z^2}=3$ maxima have $d_{yz}$ and $d_{z^2}$ orbital characters, respectively. They reside in a very narrow band formed by the weakly interacting $\alpha$ YSR states of the Fe atoms (\autoref{fig2}), which explains the low dispersion of this band. On the contrary, wide bands are formed by the strong hybridization of the $\beta$ YSR states, i.e. a $d_{x^2-y^2}$ YSR band (between -0.2 meV and +1.1 meV) having high intensities on both sides along the longitudinal axis of the chain and a $d_{xz}$ YSR band (between -0.8 meV and 0 meV) characterized by high intensities between the atoms of the chain (\autoref{fig4}a,b). In order to deduce the dispersions of these YSR bands from the theoretical calculations we applied the same 1D-FFT method as in the experiment (see also Ref. \citeonline{Schneider2021a}), and averaged over chains containing 9, 11, 13, 14, 17 and 19 Fe atoms (Supplementary Figure 7). The result is plotted in \autoref{fig4}c and can be compared to the experimental dispersion in \autoref{fig3}c. The most characteristic, broad bands are the $d_{x^2-y^2}$ and $d_{xz}$ YSR bands between -0.2 meV and 1.1 meV and between -0.8 and 0 meV, respectively. While the energy range of the $d_{xz}$ YSR band agrees reasonably well with that of the experimental $\beta$ band, the $d_{x^2-y^2}$ YSR band is probably not detected significantly in the experimental data (\autoref{fig3}c). The latter might be explained by the small intensity of the experimental $+\beta$ state (\autoref{fig1}d) which is not reproduced by the calculations (\autoref{fig2}a). Most importantly, the theoretical study confirms the lack of a detectable minigap at $E_\mathrm{F}$ in the YSR bands.
\newline
\subsection*{Minigap and end states in spin spiral Fe chains on Au monolayer}
At first sight, the missing minigap seems surprising. For similar ferromagnetic chains on the lighter substrates Nb(110) and Ta(110) there are already clear indications for the openings of topological minigaps \cite{Beck2022b, Schneider2021a}. It is widely accepted that topological minigaps hosting MBSs can open in the quasiparticle spectrum of one-dimensional helical spin systems being proximity-coupled to a conventional $s$-wave superconductor \cite{NadjPerge2013,Pientka2013,Klinovaja2013,Kim2018}. For ferromagnetic chains, this phenomenon has been attributed to a Rashba-type SOC induced by the substrate\cite{Li2014}, which is equivalent to a spin spiral structure without SOC in a single-band tight-binding model\cite{Braunecker2010}. As outlined in the introduction above, the heavier material Au is well known to exhibit large SOC\cite{Lashell1996, Henk2004}. However, as our experiments and calculations show, it obviously does not induce a spin-spiral state in the Fe chain, and likewise does not induce a SOC of sufficient strength in the YSR bands of the ferromagnetic chain to open a detectable minigap. In order to trace whether we can still force the system into a state with a large topological gap $\varDelta_\textrm{ind}$ just by artificially imposing a suitable non-collinear spin state onto the chain, we performed calculations  for the same chains as before, but now imposing a helical spin spiral state (\autoref{fig5}, Supplementary Note 8, and Supplementary Figure 8). The configuration of the spin spiral was such that the first Fe site had its spin pointing along the positive $z$ direction and then each spin is rotated by 90$^\circ$ around the chain axis when moving along the chain (\autoref{fig5}a). Indeed, there are two significant features which emerge in the LDOS of the spin-spiral chain with 19 iron atoms (\autoref{fig5}b), which were absent in the LDOS of the ferromagnetic chain (\autoref{fig4}a). First, a minigap at $E_\mathrm{F}$ opens up between $-0.22$ meV and $+0.22$ meV. Second, inside this minigap a single state can be observed at $E_\mathrm{F}$ with a pronounced intensity localized at the ends of the chain. This state has an electron-hole ratio of 1 and is robust against the variation of the chain length from 9 to 19 atoms as illustrated in \autoref{fig5}b and Supplementary Figure 8. The strongly different spatial LDOS distribution of the zero-energy state compared to that of some exemplary higher-energy states is further illustrated in \autoref{fig5}c. The former is localized over a few atoms at the two ends of the chain, while the latter states outside the minigap are extended along the whole chain. It should be mentioned that all these states, both the zero-energy one as well as those outside of the minigap show the same orbital character, indicating that the minigap emerges from the $d_{x^2-y^2}$ YSR states of the ferromagnetic chain. The induced minigap of $2\varDelta_\textrm{ind}=\SI{0.44}{\milli \electronvolt}$ width and the narrow spectral weight around $E_\mathrm{F}$ stemming from the zero-energy end states are also clearly visible in the dispersion of the scattering wave vectors deduced from the averaged 1D-FFTs of the LDOSs of chains of different lengths (\autoref{fig5}d). Thus, the calculations show evidence for the formation of a topological, most probably $p$-wave-like, minigap which hosts a MBS, if the Fe chain on Au(111) is forced into a helical spin spiral state, indicating that the absence of the non-collinear ground state is the limiting factor of this experimental system.
\subsection*{Conclusions and outlook}
\begin{figure}[H]
 	\centering
     \includegraphics[width=1\textwidth]{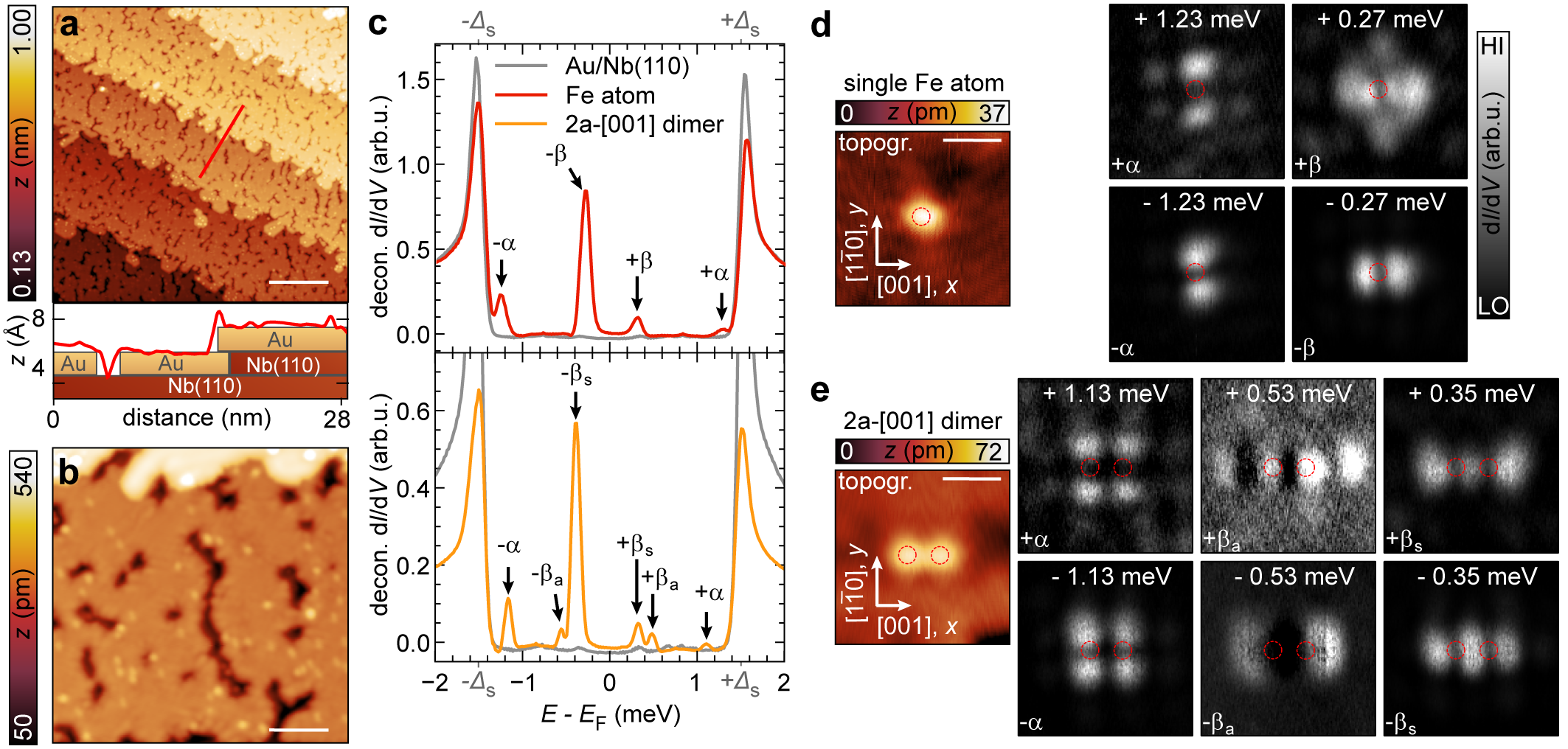}
      \caption{\label{fig1}\textbf{Measured YSR states of Fe atoms and dimers on monolayer Au on Nb(110).}\\\textbf{a}, STM image of an ultrathin film of Au on Nb(110) with an approximate coverage of 1 ML Au and additionally deposited Fe atoms. An extracted line profile along the red line is displayed in the bottom panel (red curve). Rectangles show the surface composition underneath the line profile. The white scale bar has a length of $\SI{20}{\nano \meter}$ ($\vbias=\SI{50}{\milli \volt}$ and $I=\SI{100}{\pico \ampere}$). \textbf{b}, STM image of randomly distributed Fe atoms on 1 ML Au (bottom) and 2 ML Au (top). The white scale bar has a length of $\SI{4}{\nano \meter}$ ($\vbias=\SI{6}{\milli \volt}$ and $I=\SI{3}{\nano \ampere}$). \textbf{c}, Deconvoluted \didu~spectra measured on the Au/Nb(110) substrate (gray), a single Fe atom (red, top panel), and the center of a dimer of Fe atoms spaced by $2a$ along the $[001]$ direction (orange, bottom panel). Black arrows and Greek letters label YSR states. Gray ticks mark the position of the superconducting energy gap of the sample $\varDelta_{\textrm{s}}=\SI{1.50}{\milli \electronvolt}$ as determined in Supplementary Note 2 ($\vstab=\SI{6}{\milli \volt}$, $\istab=\SI{1}{\nano \ampere}$ and $\vmod=\SI{20}{\micro \volt}$). \textbf{d} and \textbf{e}, STM images and constant-contour maps of a single Fe atom (\textbf{d}) and a Fe dimer (\textbf{e}) spaced by $2a$ along the $[001]$ direction. Constant-contour maps were obtained for every energy for which we identified a peak in the corresponding spectrum of \textbf{c} as indicated by the corresponding Greek letters. White arrows indicate crystallographic directions, red dashed circles depict the positions of the Fe atoms as determined from the topographies, and white scale bars represent a length of $\SI{1}{\nano \meter}$ ($\vbias=\SI{6}{\milli \volt}$ and $I=\SI{1}{\nano \ampere}$).}
 \end{figure}

\begin{figure}[H]
	\includegraphics[width=1\textwidth]{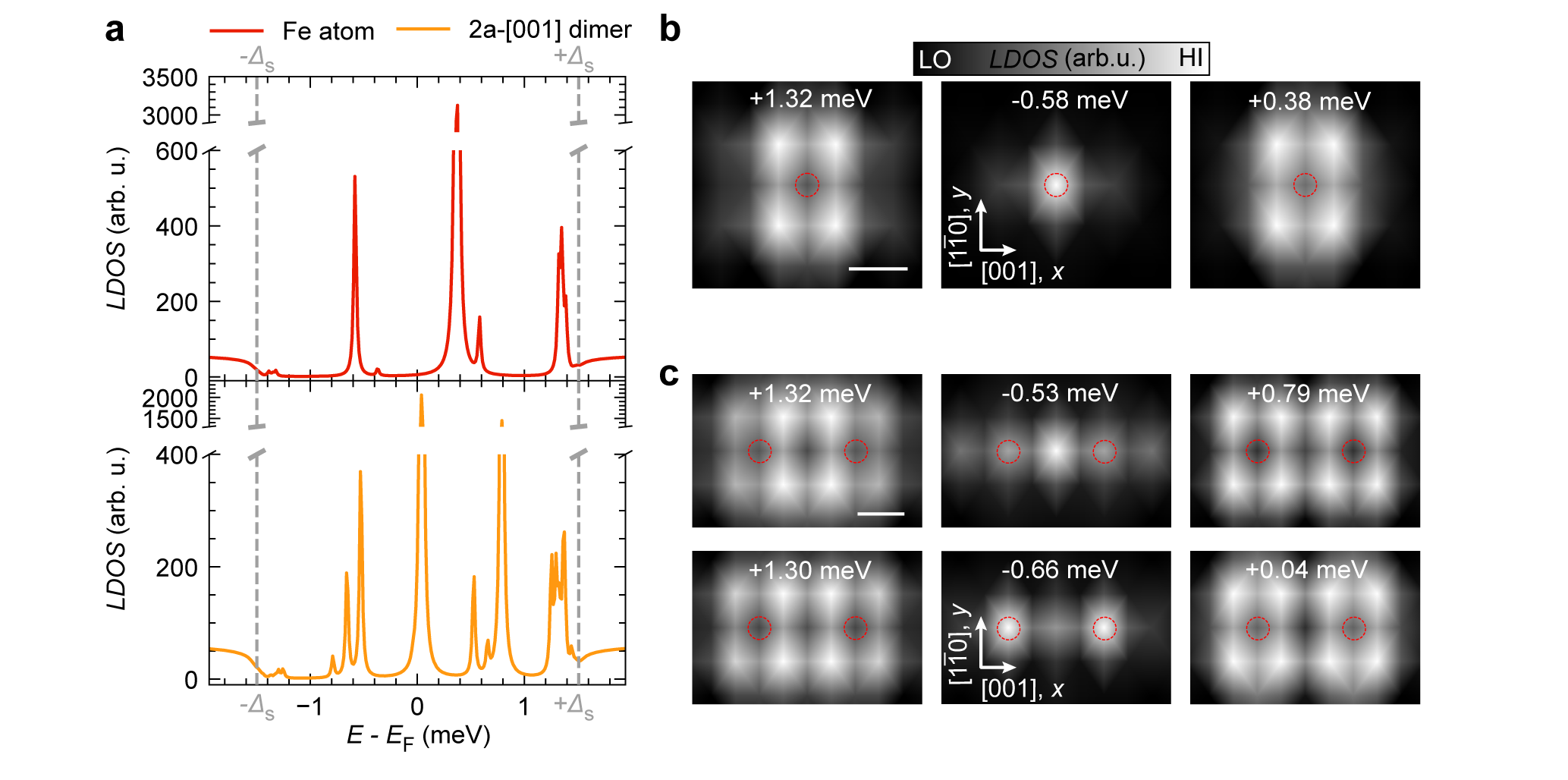} 
 	\caption{\label{fig2} \textbf{Calculated YSR states of Fe atoms and dimers on monolayer Au on Nb(110).}\\\textbf{a,} Electron component of the LDOS of the single Fe atom (top panel) and the ferromagnetic $2a-[001]$ dimer (bottom panel). Gray dashed vertical lines indicate the superconducting gap of the substrate $\varDelta_{\mathrm{s}}$. \textbf{b}, Spatial distributions of the three YSR peaks of the Fe atom with the highest intensities visible in the top panel of \textbf{a}.  \textbf{c}, Spatial distributions of the six YSR peaks of the FM $2a-[001]$ dimer with the highest intensities visible in the bottom panel of \textbf{a}. The energies are indicated in the bottom of the panels of \textbf{b} and \textbf{c}. Red circles indicate the positions of Fe atoms and the white scale bars correspond to a distance of $a$.}
 \end{figure}

\begin{figure}[H]
	\includegraphics[width=1\textwidth]{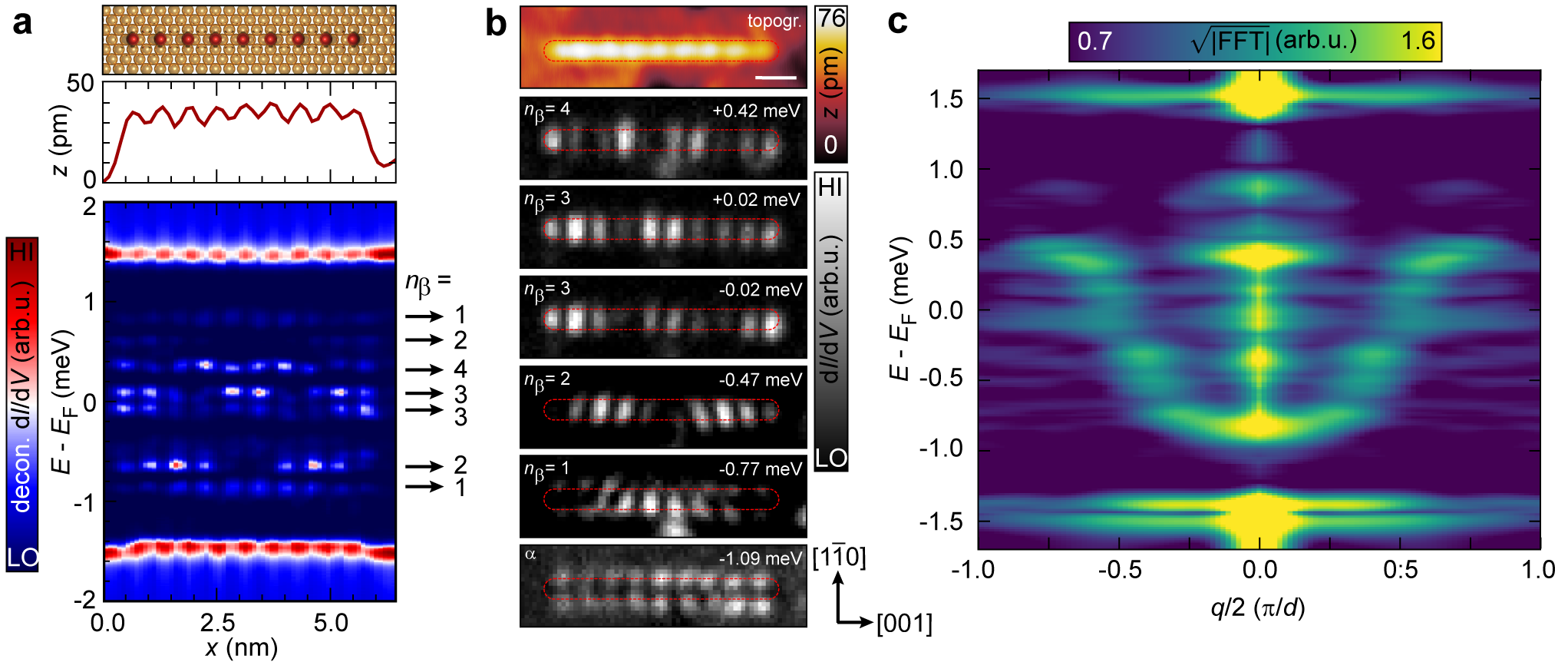} 
 	\caption{\label{fig3}  \textbf{Measured dispersion of BdG quasiparticles in Fe chains on Au monolayer.}\\\textbf{a}, Deconvoluted \didu~line profile (bottom panel) and corresponding topographic line profile (middle panel) measured along the longitudinal axis of a $\mathrm{Fe}_9$ $2a-[001]$ chain, as illustrated in the top panel. Black arrows mark the energies in the bottom panel at which $n_\beta$ maxima as indicated are observed along the chain. The subscript of this label refers to the orbital origin of this state ($\vstab=\SI{6}{\milli \volt}$, $\istab=\SI{1}{\nano \ampere}$, $\vmod=\SI{20}{\micro \volt}$). \textbf{b}, The top panel shows an STM image of a $\mathrm{Fe}_9$ $2a-[001]$ chain and the lower panels are $\mathrm{d}I/\mathrm{d}V$ maps of this chain, obtained at energies indicated in the top right corner of each panel. The maps are labeled by $n_\beta$ in a similar fashion as the states in \textbf{a}. The red lines mark the spatial extent of the chain in the STM image. The white scale bar represents a length of \SI{1}{\nano \meter} ($\vstab=\SI{-6}{\milli \volt}$, $\istab=\SI{1}{\nano \ampere}$, $\vmod=\SI{20}{\micro \volt}$). \textbf{c}, Averaged energy-wise 1D-FFT obtained from \didu~line profiles of Fe$_N$ $2a-[001]$ chains with lengths $N$ ranging from seven to fourteen atoms. Prior to the 1D-FFT, the spectra were deconvoluted. All \didu~line profiles were obtained with the following parameters: $\vstab=\SI{6}{\milli \volt}$, $\istab=\SI{1}{\nano \ampere}$, $\vmod=\SI{20}{\micro \volt}$.}
 \end{figure}

\begin{figure}[H]
\centering
	\includegraphics[width=1\textwidth]{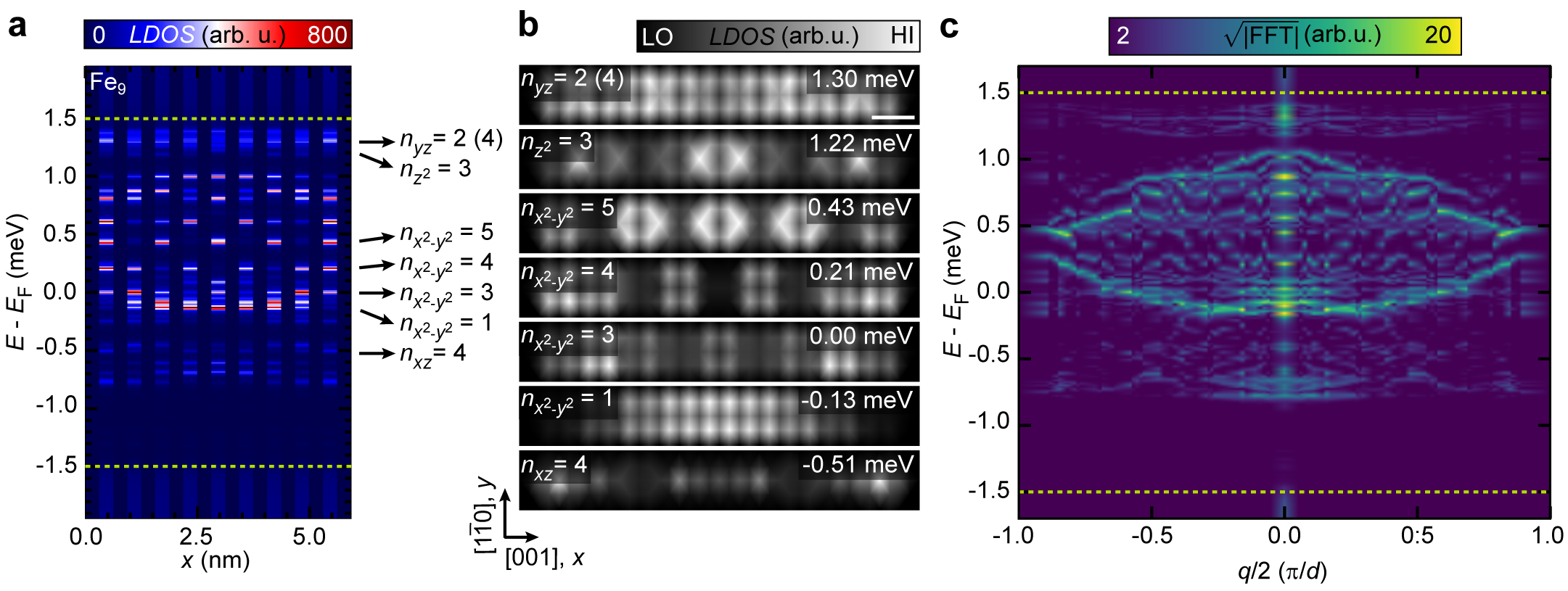} 
 	\caption{\label{fig4} \textbf{Calculated dispersion of BdG quasiparticles in ferromagnetic Fe chains on Au monolayer.}\\\textbf{a}, Electron component of the LDOS extracted along the longitudinal axis of a ferromagnetic Fe$_9$ $2a-[001]$ chain, calculated on the Fe and the vacuum sites in between. \textbf{b}, Spatial distributions of the LDOS evaluated in the first vacuum layer above the chain.  Energies are indicated in the top right corners of each panel, while the numbers of maxima and the orbital origins of the states are indicated in the top left corners. They correspond to the states marked by the same arrows and numbers in \textbf{a}. The $n_{yz}$ state has two dominant Fourier components $n_{yz}=2$ and $n_{yz}=4$, where the former is dominating. The white scale bar has a length of $2a$. \textbf{c}, Dispersion of scattering wave vectors extracted from the calculated LDOSs of ferromagnetically coupled Fe$_N$ $2a-[001]$ chains and averaged for lengths $N$ of 9, 11, 13, 14, 17 and 19 atoms (Supplementary Figure 7). The green dashed lines in panels \textbf{a} and \textbf{c} indicate the energy gap of the superconducting substrate.}
 \end{figure}

\begin{figure}[H]
\centering
	\includegraphics[width=1\textwidth]{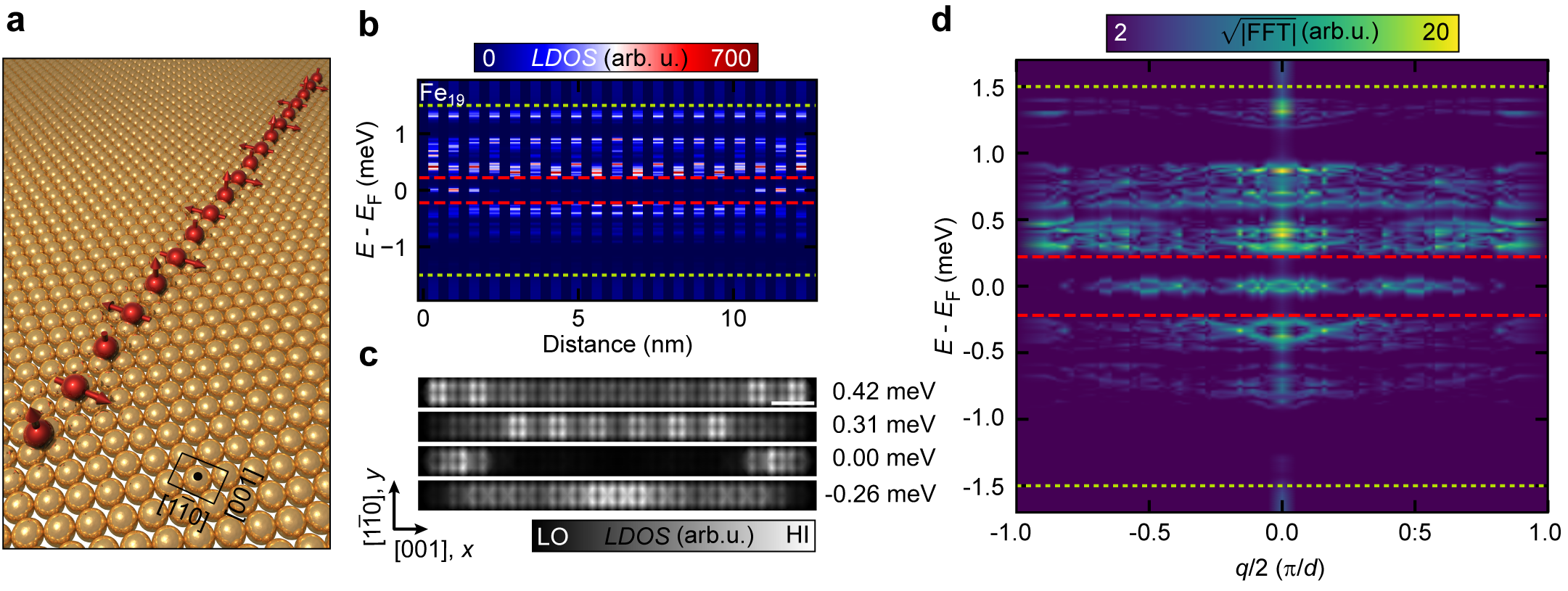} 
 	\caption{\label{fig5}\textbf{Minigap and MBS enforced by helical spin spirals in Fe chains on Au monolayer.} \\\textbf{a}, Illustration of the helical spin spiral state with a rotation angle of 90$^\circ$ of a chain containing 19 Fe atoms. \textbf{b}, Electron component of the LDOS extracted along the longitudinal axis of a Fe$_{19}$ $2a-[001]$ chain in the helical spin spiral state shown in panel \textbf{a}.  \textbf{c}, The spatial distributions of the LDOS evaluated in the first vacuum layer above the Fe$_{19}$ $2a-[001]$ chain at the energies indicated in the top right corner of each panel. \textbf{d}, Dispersion of scattering wave vectors averaged from the calculated LDOSs of the Fe$_N$ $2a-[001]$ chains including $N=$ 9, 11, 13, 14, 17 and 19 Fe atoms (Supplementary Figure 8). The green and red dashed lines in \textbf{b} and \textbf{d} indicate the substrate gap and the minigap, respectively.}
 \end{figure}

In summary, our combined experimental and theoretical investigation shows that in contrast to what might be suggested by simplified tight-binding models\cite{Li2014, Braunecker2010}, a strong substrate SOC alone generally is not a sufficient condition for the opening of a topological minigap in a ferromagnetic chain in contact to an $s$-wave superconductor, since the SOC has to exist in the lowest-energy YSR band.
In fact, first-principles calculations of the magnetic interaction parameters in ultrathin film systems have demonstrated that also the connection between the formation of a spin-spiral state and SOC is considerably more complicated. In particular, the DMI preferring a non-collinear spin alignment is typically weak when a $3d$ transition metal is deposited on a Au surface compared to other $5d$ substrates\cite{Simon2014,Belabbes2016,Simon2018}, which may be tentatively attributed to the fully occupied $5d$ band of Au having a reduced effect on the DMI. Proximity to a Au layer is known to give rise to strong Heisenberg exchange interactions and anisotropy\cite{Szunyogh1995} in the magnetic layer instead, both of which prefer a collinear spin alignment and the latter being induced by the SOC. Our results indicate that, similarly to the competition between DMI and anisotropy terms in the formation of non-collinear spin structures, the role of SOC may be more complex for inducing topological superconductivity in the YSR bands of ferromagnetic spin chains.
\newline
Our study proves that it is experimentally possible to grow proximitized ultrathin heavy metal layers on a superconductor with a large $T_{\textrm{c}}$ that can be used as a substrate for the deposition of transition metal atoms and to construct defect-free one-dimensional structures with an excellent quality, enabling the tailoring of YSR bands. Further, we presented an \textit{ab-initio} method that accurately reproduces the main LDOS features observed in the experiments. Our work thus demonstrates the theoretical feasibility of an \textit{ab-initio} screening of other combinations of transition metal chains on heavy metal thin films on bulk superconductors in order to find the optimal conditions for the opening of a large topological minigap.
\begin{methods}

\subsection*{STM and STS measurements}
The experiments were performed in a custom home-built ultra-high vacuum system, equipped with an STM setup, which was operated at a temperature of $\SI{320}{\milli \kelvin}$\cite{Wiebe2004}. 
STM images were obtained by applying a bias voltage $V_{\mathrm{ bias}}$ to the sample upon which the tip-sample distance is controlled by a feedback loop such that a constant current $I$ is achieved.
\didu~spectra were obtained in open feedback mode after stabilizing the tip at $\vstab=\SI{6}{\milli \volt}$ and $\istab=\SI{1}{\nano \ampere}$ using a standard lock-in technique with an AC voltage  $\vmod=\SI{20}{\micro \volt}$ (rms value) of frequency $f_{\mathrm{mod}}=\SI{4142}{\hertz}$ added to the ramped $\vbias$. If other stabilization parameters were used for a particular measurement, it is indicated in the respective figure caption.
\didu~maps were obtained by measuring \didu~spectra on a predefined spatial grid, which was positioned over the structure of interest, and selecting a slice at a given voltage. Typical measurement parameters are the same as for individual \didu~spectra.
\didu~line profiles are measured similarly to \didu~maps, with the exception that the spatial grid is one-dimensional.
Constant-contour maps were obtained by repeated scanning of individual lines of STM images. First, each line is measured as it would be the case in a regular STM image. The $z$-signal of this sweep is saved. Then, the feedback is turned off, the bias voltage $\vbias$ is set to a predefined value of interest, and for the next sweep on the same scan line the \didu signal is measured while restoring the previously recorded $z$-signal. 
\newline
A mechanically sharpened and in-situ flashed (\SI{50}{\watt}) bulk Nb tip was used for all measurements. While the usage of a superconducting tip is a crucial factor for obtaining a very good energy resolution, it has the downside that the \didu~spectra are convolutions of the tip and sample DOSs. However, we can determine the superconducting gaps of the tip and the sample, and deconvolute the \didu~spectra. This process is described in Supplementary Note 2 and is performed for every spectrum in the main manuscript.

\subsection*{Sample preparation}
A Nb(110) single crystal with a purity of $\SI{99.99}{\percent}$ was transferred into the ultra-high vacuum chamber. The sample was cleaned by cycles of Ar ion sputtering and flashes up to $\SI{2400}{ \celsius}$, which results in a clean surface with only few oxygen impurities remaining\cite{Odobesko2019}. We established flashing parameters which clean the surface of oxygen, and checked the results by STM. Once this cleaning procedure was reproducible with the given parameters, we evaporated Au from an e-beam evaporator (EFM3 by FOCUS GmbH) equipped with a Au rod ($\SI{99.99}{\percent}$ purity). Following this procedure, we achieved flat and spatially extended films (\autoref{fig1}a).
\newline
Fe was evaporated onto the surface from a carefully outgassed Fe rod using a second e-beam evaporator while keeping the sample temperature below $T=\SI{10}{\kelvin}$ to avoid clustering and diffusion and thus achieve a random distribution of single Fe adatoms (\autoref{fig1}b). From Supplementary Note 1, we conclude that the Au film grows pseudomorphically and that the Fe atoms are adsorbed in the fourfold coordinated hollow sites in the center of four Au atoms. This is further supported by the similarity of the spatial distributions of the YSR states of the Fe/Au/Nb(110) system compared to the Mn/Nb(110) system\cite{Beck2021}.
\newline
STM tip-induced atom manipulation\cite{Stroscio2004, Eigler1990} is used to position individual Fe atoms and construct artificial structures, such as dimers and chains. The structures built in this study have sufficient interatomic spacing to unambiguously identify the positions of the individual atoms forming the structure using STM images. We restrict the investigations here to Fe atoms positioned on the first ML of Au. Fe atoms on the first and second  ML can easily be distinguished by their apparent height. In the top part of \autoref{fig1}b one can see that an Fe atom on the second ML appears as a bright spherical protrusion, while an Fe atom on the first ML is more shallow and has a relatively irregular shape. Thus, we can be sure that all of the experiments were carried out on the first ML.

\subsection*{First-principles calculations} The calculations were performed in terms of the Screened Korringa-Kohn-Rostoker method (SKKR), based on a fully relativistic Green's function formalism by solving the Dirac equation for the normal state\cite{Szunyogh1995} and the Dirac-Bogoliubov-de Gennes (DBdG) equation for the superconducting state within multiple scattering theory (MST)\cite{csire2018relativistic, Nyari2021}. The impurities are included within an embedding scheme\cite{Lazarovits2002}, being an efficient method to address the electronic and magnetic properties or the in-gap spectra of real-space atomic structures without introducing a supercell. The system consists of seven atomic layers of Nb, a single atomic layer of Au and four atomic layers of vacuum between semi-infinite bulk Nb and semi-infinite vacuum. The Fe adatoms are placed in the hollow position in the vacuum above the Au layer and relaxed towards the surface by 21\%, while the top Au layer is also relaxed inwards by 2\%. 
The relaxations are obtained from total-energy minimization in a VASP\cite{Kresse1996,Kresse1996a,Hafner2008} calculation for a single Fe adatom and are used for the dimer and all the chains. For the potentials we employ the atomic sphere approximation (ASA), the normal state is calculated self-consistently in the local density approximation (LDA) as parametrized by Vosko \textit{et al.}\cite{Vosko1980} The partial waves within MST are treated with an angular momentum cutoff of $\ell_\mathrm{max}=2$. In the self-consistent normal state calculations we used a Brillouin zone (BZ) integration with 253 $\mathbf{k}$ points in the irreducible wedge of the BZ and a semicircular energy contour on the upper complex half plane with 16 points for energy integration. In order to take into account charge relaxation around the magnetic sites, the single impurity and the $2a-[001]$ dimer are calculated by including a neighborhood containing 48 and 84 atomic sites, respectively, corresponding to a spherical radius of $r=1.66~a$. The atomic chains are calculated with a somewhat smaller neighborhood corresponding to 2 atomic shells or a spherical radius of $r=1.01~a$ around the Fe atoms. This way our largest atomic cluster in the calculation with 19 Fe chain atoms contained 339 atomic sites.
After having obtained the self-consistent potentials in the normal state, the superconducting state is simulated within single-shot calculations by solving the DBdG equation with the experimental band gap used as the pairing potential in the Nb layers\cite{Nyari2021}. In the case of the single impurity and the dimer, the BZ integration for the host Green's function is performed by using the same $\mathbf{k}$ mesh as for the normal state, but in order to achieve convergence for the chains we had to increase the number of $\mathbf{k}$ points up to 1891 in the irreducible wedge of the BZ. A sufficient energy resolution of the LDOS in the superconducting gap is acquired by considering 301 energy points between $\pm 1.95$~meV with an imaginary part of 13.6~$\mu$eV related to the smearing of the resulting LDOS. Both the electron and the hole components of the LDOS are calculated, but in this paper we present only the electron part leading to the asymmetry of the spectrum as also seen in the experiments. Due to the ASA used in our method we obtain the LDOS for each atomic site of the cluster averaged inside the atomic spheres. The orbital resolution of the YSR states can be determined based on the orbital-resolved LDOS of the Fe atoms. Since the canonical $d$ orbitals hybridize due to the symmetry of the cluster and due to SOC, we assign the labels based on the orbital which has the largest contribution to the given peak in the LDOS. In addition, in order to mimic the constant-contour maps in the experiments, we evaluate the spatial distribution of the LDOS. These LDOS maps are taken from the first vacuum layer above the surface in which the magnetic atoms are embedded, reflecting the orbital characteristics obtained from the resolution of the LDOS of the Fe atoms. 
In order to better reproduce the experimental constant-contour maps taken from the vacuum region, the LDOS of the magnetic sites are replaced by the average LDOS over the two vacuum sites (empty spheres) closest to them in the layer above. To get a continuous picture for the LDOS maps we applied an interpolation\cite{gouraud1971} scheme on the data calculated as described above.

\subsection*{Data availability}
The authors declare that all relevant data are included in the paper and its Supplementary Information files.

\end{methods}



\begin{addendum}
 \item P.B., R.W., and J.W. gratefully acknowledge funding by the Deutsche Forschungsgemeinschaft (DFG, German Research Foundation) – SFB-925 – project 170620586. L.S., R.W., and J.W. gratefully acknowledge funding by the Cluster of Excellence 'Advanced Imaging of Matter' (EXC 2056 - project ID 390715994) of the DFG. R.W. gratefully acknowledges funding of the European Union via the ERC Advanced Grant ADMIRE (project no. 786020).
 B.Ny., L.R., A.L., K.P., L.Sz. and B.U. acknowledge financial support by the National Research, Development, and Innovation Office (NRDI) of Hungary under Project Nos. FK124100 and K131938. B.Ny. and L.Sz. acknowledge support by the Ministry for Innovation and Technology and the NRDI Office within the Quantum Information National Laboratory of Hungary. B.Ny. and K.P. acknowledge the support by the ÚNKP-21-3 and the ÚNKP-21-5 New National Excellence Program of the Ministry for Innovation and Technology from the source of the National Research, Development and Innovation Fund. K.P. acknowledges the János Bolyai Research Scholarship of the Hungarian Academy of Sciences. 

 \item[Author contributions] P.B., L.S. and J.W. conceived the experiments. P.B. and L.S. performed the measurements and P.B. analyzed the experimental data together with J.W.. L.R. performed the VASP calculations. B.N. performed the SKKR calculations and discussed the data with L.R., A.L., K.P., L.Sz. and B.U., and A.L. also contributed to the spin model calculations. P.B. and B.N. prepared the figures and wrote the first version of the manuscript. L.S., L.R., A.L., K.P., L.Sz., B.U., J.W. and R.W. contributed to the discussions and the finalization of the manuscript.

 \item[Competing Interests] The authors declare no competing interests.

 \item[Correspondence] Correspondence and requests for materials
should be addressed to J. Wiebe~(email: jwiebe@physnet.uni-hamburg.de).
\end{addendum}

\newpage

\newpage

\end{document}